\documentclass[12pt]{iopart}
\usepackage{iopams}

\def\Journal#1#2#3#4{{#4} {\it #1} {\bf #2}, #3 }

\def\e{\epsilon}

\def\s{\sigma}
\def\t{\theta}

\begin{document}

\title{Anti-Newtonian universes do not exist}

\author{Lode Wylleman}

\address{Faculty of Applied Sciences TW16, Gent University, Galglaan 2, 9000 Gent, Belgium}

\begin{abstract}
In a paper by Maartens, Lesame and Ellis (1998) it was shown that
irrotational dust solutions with vanishing electric part of the Weyl
tensor are subject to severe integrability conditions and it was
conjectured that the only such solutions are FLRW spacetimes. In
their analysis the possibility of a cosmological constant $\Lambda$
was omitted. The conjecture is proved, irrespective as to whether
$\Lambda$ is zero or not, and qualitative differences with the case
of vanishing \emph{magnetic} Weyl curvature are pointed out.
\end{abstract}


\section{Introduction}

Irrotational dust (ID) spacetimes are perfect fluid solutions of the
general relativistic field equations characterized by vanishing
pressure ($p=0$) and vorticity vector ($\omega^a=0$). They serve as
potential models for the late universe
\cite{EllisNel,MaartensMatravers} and gravitational collapse
\cite{Bertschinger}. In an initial-value problem formulation for
such models, the Ricci-identity for the fluid 4-velocity and the 
Bianchi equations (incorporating the field equations via the 
Ricci tensor) may be covariantly
split into propagation and constraint equations \cite{Ellis}. In a
streamlined setting \cite{Maartens1} Maartens showed  that these
constraint equations are consistent with each other (i.e., not
overdetermined on an initial hypersurface) and are preserved under
evolution, generically. However, when \emph{further} external
conditions are imposed their consistency is not a priori guaranteed,
and rather unexpected conclusions may come out.

In this respect, one of the most promising and natural further
restrictions one can make is to set the gravito-magnetic tensor
$H_{ab}$ equal to zero (a cosmological motivation for doing so was
given in \cite{Matarrese1}), with the remaining gravito-electric
tensor $E_{ab}\neq 0$. As the latter is the relativistic
generalization of the tidal tensor in Newtonian theory, the
corresponding `purely electric' ID models were called
\emph{Newtonian-like} in \cite{Maartens2}. One might therefore
expect to find a rich class of solutions, at least incorporating
translates of classical Newtonian models. However, in two
independent papers \cite{vanElst,Sopuerta} a non-terminating chain
${\cal C}_E$ of consistency conditions for the dynamical variables
was deduced, which made the authors conjecture that the class was
unlikely to extend beyond the known spatially homogeneous members
and the Szekeres \cite{Szekeres} subfamily, which exhausts Petrov
type D \cite{Barnes}.

The natural counterpart to the above is to set $E_{ab}$ equal to
zero, with $H_{ab}\neq 0$. These purely magnetic ID models were
studied in \cite{Maartens2} and were logically called
`anti-Newtonian'. Some parallels and differences between
Newtonian-like and anti-Newtonian universes were pointed out. In
both cases no spatial derivatives occur in the remaining propagation
equations; hence the flowlines emerging from an initial hypersurface
evolve separately from each other, and solutions may therefore be
called `silent' \cite{Maartens2}, although in the literature this
terminology is normally used and was introduced \cite{Matarrese2}
for the purely electric case only. Also, the chain ${\cal C}_E$ for
the Newtonian-like case has a direct analogue ${\cal C}_{H,1}$ in
the anti-Newtonian case. However, whereas ${\cal C}_E$ is
identically satisfied in the linearized theory (with
Friedman-Lemaitre-Robertson-Walker (FLRW) background), leading to a
\emph{linearization instability} for Newtonian-like ID models (in the sense that
there are consistent linearized solutions which are not the limit of
any consistent solutions in the full, nonlinear theory \cite{vanElst}), it
was shown by the primary integrability condition in ${\cal C}_{H,1}$
that there are no linearized anti-Newtonian ID models at all. On top
of this, a second chain ${\cal C}_{H,2}$ of integrability conditions
exists for the magnetic case, which has no analogue in the electric
case. Both facts have lead the authors of \cite{Maartens2} to
conjecture the \emph{non-existence of purely magnetic ID models},
but a proof for it within the full, nonlinear theory was not found
until now.

It is usually believed that the presence of a cosmological constant
$\Lambda$ is of minor or no importance to the qualitative outcome of
consistency analyses for particular conditions, and in this line
$\Lambda$ was omitted in \cite{vanElst,Sopuerta} and
\cite{Maartens2}. For Newtonian-like silent models however, one of
the special cases where one can show that the constraints with
$\Lambda\leq 0$ are inconsistent (namely the case where one of the
eigenvalues of the expansion tensor is 0), surprisingly turns out to
admit two new and explicitly constructable families of spatially
\emph{in}homogeneous Petrov type I metrics when $\Lambda > 0$
\cite{Vandenbergh1}. This result points out that a cosmological
constant should preferably be incorporated in consistency analyses.

In the present paper and in this respect, the
generalization to arbitrary cosmological constant $\Lambda$ of the
above Maartens \emph{et al} conjecture for purely magnetic ID
models is proved, by showing that the combined conditions of vanishing pressure
$p$, vorticity vector $\omega^a$ and gravito-electric tensor
$E_{ab}$ are only consistent for FLRW spacetimes. The structure of
the paper is as follows. In section 2.1 the 1+3
covariant system of propagation and constraint equations for general
irrotational dust is written down as in \cite{Maartens2}, with a
cosmological constant added. In section 2.2 the origin of
the two chains of integrability conditions for the purely magnetic
case is resumed and it is pointed out how a third chain is coming into play. Comparison
with the purely electric case is made in Section 2.3, and the
conjecture for Newtonian-like ID models \cite{vanElst,Sopuerta} is
properly restated in function of the sign of $\Lambda$. The actual
proof (in two steps) for the inconsistency of anti-Newtonian
universes is the content of Section 3. A natural continuation of the
analysis is pointed out in Section 4. 

\section{Magnetic versus electric ID}
\subsection{Basic setting for general ID}

Einstein's field equation with cosmological constant $\Lambda$ and
perfect fluid source term $T_{ab}=\mu u_au_b+p\,h_{ab}$ can be
covariantly written as ($8\pi G=1=c$):
\begin{eqnarray}\label{fieldeq}
        R_{ab}&=&\frac{1}{2}(\rho +3p)u_a u_b+\frac{1}{2}(\rho-p)h_{ab}-\Lambda g_{ab}.
\end{eqnarray}
It provides an algebraic definition of the Ricci tensor $R_{ab}$
in terms of the matter density $\rho$, pressure $p$ and normalized
timelike four-velocity field $u^a$ ($u^a u_a=-1$) of the perfect
fluid; the corresponding pair of complementary idempotent
operators $(-u_a\,u^b,{h_{a}}^{b}:={g_{a}}^{b}+u_a\,u^b)$ projects
along and orthogonal to $u^a$, respectively.

The introduction of the following 1+3 covariant projections and
operations (see \cite{Maartens1,Maartens3}) considerably
compactifies expressions. Herein round (square) brackets denote
symmetrization (anti-symmetrization) and  $\textrm{tr}(S)$ is written for ${S^a}_a$; for 2-tensors
$U_{ab}$ and one-forms $X_a$ the `dot-free' notation $(SU)_{ab}$
for ${S_a}^cU_{cb}$, resp.\ $(SX)_a$ for ${S_a}^b X_b$, is used
throughout the paper.

\begin{itemize}
  \item The \emph{spatial part} of a tensor ${S^{a_1\ldots a_r}}_{b_1\ldots b_s}$ is
\begin{eqnarray}\label{Sspat}
    {}^{s} {S^{a_1\ldots a_r}}_{b_1\ldots b_s}
    &=&{h^{a_1}}_{c_1}\cdots {h^{a_r}}_{c_r}{h_{b_1}}^{d_1}\cdots {h_{b_s}}^{d_s} {S^{c_1\ldots c_r}}_{d_1\ldots d_s}.
\end{eqnarray}
A tensor is called \emph{spatial} iff ${}^{s} {S^{a_1\ldots
a_r}}_{b_1\ldots b_s}={S^{a_1\ldots a_r}}_{b_1\ldots b_s}$. The spatial, symmetric and tracefree part $S_{\langle ab\rangle}$ of
a 2-tensor $S_{ab}$ is 
\begin{eqnarray}
S_{\langle ab\rangle}&=&{}^{s}S_{(ab)}-\frac{1}{3}\textrm{tr}({}^{s}S)h_{ab}.
\end{eqnarray}
  \item  The \emph{spatial permutation tensor} is $\e_{abc}=\e_{[abc]}=\eta_{abcd}\,u^d$, where
$\eta_{abcd}=\eta_{[abcd]}$ is the spacetime Levi-Civita
permutation tensor. One then defines for 2-tensors
$S_{ab},U_{ab}$:
\begin{eqnarray}
    [S,U]_a=\e_{abc}(SU)^{bc}.
\end{eqnarray}
When $S_{ab}$ and $U_{ab}$ are spatial and symmetric, one can easily
show the identity (see also $(A3) of \cite{Maartens1}$)
\begin{equation}\label{S2U}
  [S^2,U]_a = 2\,[S,\widehat{SU}]_a=\textrm{tr}(S)\,[S,U]_a-(S\,[S,U])_a,
\end{equation}
with $\widehat{SU}_{ab}=(SU)_{\langle ab \rangle}$.
In this case $[S,U]_a=0$ is equivalent to $(SU)_{ab}$ being
symmetric, which is still equivalent to $(SU-US)_{ab}=0$, whence
${S_a}^b$ and ${U_{a}}^b$ have a common eigenframe.
  \item The \emph{electric part} $E_{ab}$ and \emph{magnetic part} $H_{ab}$ of the Weyl tensor
  $C_{abcd}$ are
\begin{eqnarray}
    E_{ab}=E_{\langle
ab\rangle}=C_{acbd}\,u^c\,u^d,&\quad&H_{ab}=H_{\langle
ab\rangle}=\e_{cda}{C^{cd}}_{be}\,u^e.
\end{eqnarray}
  \item Let $\nabla$ be the spacetime covariant derivative attached
to the unique metric connection. The \emph{(covariant) time
derivative} $T$ and the \emph{(covariant) spatial derivative} $D$
are defined by:
\begin{eqnarray}
  \label{defT}{T(S)^{a_1\ldots a_r}}_{b_1\ldots b_s}&=&u^c\nabla_c {S^{a_1\ldots a_r}}_{b_1\ldots b_s},\\
    \label{defD} D_c\, {S^{a_1\ldots a_r}}_{b_1\ldots b_s}&=&{}^{s}\nabla_c {S^{a_1\ldots a_r}}_{b_1\ldots b_s}.
\end{eqnarray}
Normally the notation ${\dot{S}^{a_1\ldots a_r}}$$_{b_1\ldots
b_s}$ is used instead of ${T(S)^{a_1\ldots a_r}}_{b_1\ldots b_s}$,
but the latter emphasizes on the action of $T$ as a (degree 0
derivative) operator and this is more convenient for what follows.
Definition (\ref{defD}) tells that $D$ maps each $(r,s)$-tensor
${S^{a_1\ldots a_r}}_{b_1\ldots b_s}$ to the spatial part of the
$(r,s+1)$-tensor $\nabla_c {S^{a_1\ldots a_r}}_{b_1\ldots b_s}$.
It gives rise to the covariant spatial divergence (div) and curl
operator, acting on vectors $V^a$ and 2-tensors $S_{ab}$ as:
\begin{eqnarray}
    \textrm{div}\, V:=D_a V^a,&\quad&{\textrm{curl}\, V}_a:=\e_{abc}D^b\,V^c\\
    {\textrm{div}\, S}_a:=D^b S_{ab},&\quad&{\textrm{curl}\, S}_{ab}:=\e_{cd(a}D^c {S_{b)}}^d;
\end{eqnarray}
For spatial 2-tensors $S_{ab},U_{ab}$ one has the identity
\begin{eqnarray}\label{divSU}
    \textrm{div}[S,U]&=&\textrm{tr}(U\,\textrm{curl}S)-\textrm{tr}(S\,\textrm{curl}U).
 \end{eqnarray}
  \item The splitting of $\nabla u$ into different kinematical quantities \cite{Ellis}:
\begin{eqnarray}
    u_{a;b}&=&D_b u_a-\dot{u}_au_b\nonumber\\
    \label{uab}&:=&\s_{ab}+\frac{1}{3}\theta h_{ab}+\e_{abc}\omega^c-\dot{u}_au_b.
\end{eqnarray}
Herein $\s_{ab}=u_{\langle a;b\rangle}$ is the shear tensor,
$\theta={u^a}_{;a}$ the volume-expansion scalar,
$\omega^a=\frac{1}{2}{\e^{abc}}\,u_{b;c}$ the vorticity vector, and
$\dot{u}^a$ the acceleration vector, which are all spatial.
\end{itemize}

For this introduction, the abstract index notation \cite{Wald} was
most transparent, but an index-free notation turns out
to be practical for calculation purposes. In particular,
$E_{ab},H_{ab},\s_{ab},h_{ab},\omega^a,\dot{u}^a$ are denoted
$E,H,\s,\omega,\dot{u}$ from now on. Moreover, since all appearing
1-forms and 2-tensors will be spatial the symbol $h$ will be dropped
everywhere. The notation for $S_{\langle ab\rangle}$ becomes
$\hat{S}$ (cf.\ \cite{Ferrando}), but when $S$ is symmetric this is written
as $S-\frac{1}{3}\textrm{tr}S$ for computational
convenience. In this notation, the $\mathbb{R}$-linear operators $T$
and $D$ satisfy in particular the Leibniz rules
\begin{eqnarray}\label{TLeib}
T(XY)=T(X)\,Y+X\,T(Y),\quad D(fg)=(Df)\,g   +f\,Dg
\end{eqnarray}
for 2-tensors or scalar functions $X$, 2-tensors, 1-forms or scalar
functions $Y$, and scalar functions $f,g$. In the present paper we
deal with \emph{non-vacuum irrotational dust (ID)}, i.e.\ solutions
of (\ref{fieldeq}) for which
\begin{eqnarray}\label{nonvacID}
    p=0,\quad \rho\neq 0,\quad \omega=0,\quad \dot{u}=0,
\end{eqnarray}
where $\dot{u}=0$ is a mere consequence of $p=0,\rho\neq 0$ and
the second contracted Bianchi identity. With $\dot{u}=0$ one has
\begin{eqnarray}\label{TSU}
    T[S,U]=[T(S),U]+[S,T(U)].
\end{eqnarray}
A commutator rule for $T$ and $D$, written in
appropriate form and for the action on scalar functions $f$, is also needed. With
$\omega=\dot{u}=0$ it reads \cite{Maartens1}:
\begin{equation}\label{commTD}
    T(Df)=D(T(f))-(\s+\frac{1}{3}\theta) Df.
\end{equation}

\vspace{.5cm}

For perfect fluids in general the Ricci identity for $u$ and the
Bianchi-identities, hereby substituting (\ref{fieldeq}) for the
Ricci tensor, are basic equations to be satisfied. They may be
covariantly split into a set of propagation equations (involving the
covariant time derivative) and a set of constraint equations
\cite{Ellis,Maartens1}. In the case of non-vacuum ID
(\ref{nonvacID}) the sets of non-identical equations read:
\begin{itemize}
    \item Propagation equations:
\begin{eqnarray}
    \label{rhodot2}T(\rho)=-\rho\theta,\\
    \label{thetadot2}T(\theta)=-\frac{1}{3}\t^2-\textrm{tr}(\sigma^2)-\frac{1}{2}\rho+\Lambda,\\
    \label{sdot2}T(\s)
    =-\frac{2}{3}\t\s-\sigma^2+\frac{1}{3}\textrm{tr}(\s^2)-E\\
    \label{Edot}T(E)-\textrm{curl}H=-\theta E+3\widehat{\s E}
    -\frac{1}{2}\rho\s,\\
    \label{Hdot}T(H)+\textrm{curl}E=-\theta H+3\widehat{\s H}
\end{eqnarray}
\item Constraint equations:
\begin{eqnarray}
    \label{C1}{{\cal C}^1}&\equiv&\textrm{div}\s-\frac{2}{3}D\t=0,\\
    \label{C2}{{\cal C}^2}&\equiv&\textrm{curl}\s-H=0,\\
    \label{C3}{{\cal C}^3}&\equiv&\textrm{div}E-\frac{1}{3}D\rho-[\s,H]=0,\\
    \label{C4}{{\cal C}^4}&\equiv&\textrm{div}H+[\s,E]=0.
\end{eqnarray}
\end{itemize}

\vspace{.5cm}

On considering (\ref{sdot2}) and (\ref{C2}), (\ref{Edot}),
(\ref{C1}) and (\ref{C3}), one sees
that within the class of non-vacuum ID spacetimes, the FLRW case
\begin{eqnarray}\label{FLRW}
     D\rho=D\theta=0,&\quad&\s=E=H=0.
\end{eqnarray}
is covariantly characterized by vanishing shear
tensor ($\s=0$) or, equivalently, by Petrov type 0 ($E=H=0$).

In \cite{Maartens1} it was shown that the time derivatives of each
of the $C^i,i=1\,..\,4$, are differential combinations of the $C^j$
themselves, such that (\ref{C1})-(\ref{C4}) evolves consistently and
doesn't give rise to new constraints, generically. It was also shown
that actually, ${\cal C}^4$ is a differential combination of ${\cal
C}^1$ and ${\cal C}^2$:
\begin{eqnarray*}\label{C412}
  {\cal C}^4 &=& \frac{1}{2}\textrm{curl}\,{\cal C}^1-\textrm{div}\,{\cal C}^2.
\end{eqnarray*}
However, if extra covariant conditions are imposed, new
constraints may be generated and their consistency should be
investigated.

\subsection{Purely magnetic ID}

In this respect, the constraint $E=0$ was studied as the natural
counterpart for $H=0$ in \cite{Maartens2}. The remaining variables
are the scalars $\rho,\theta$ and the tensorial quantities $\s,H$.
As put forward in the introduction, one sees that no spatial
derivatives occur in the propagation equations
(\ref{rhodot2})-(\ref{sdot2}) and (\ref{Hdot}), such that the purely
magnetic ID models are `silent'. Actually, with $E=0$,
(\ref{rhodot2})-(\ref{sdot2}) forms a closed subsystem determining
the evolution of the matter variables $\rho,\theta$ and $\s$, from
which the evolution equation (\ref{Hdot}) for the gravito-magnetic
field $H$ is decoupled. The latter equation may also be seen as
arising from the curl of the algebraic constraint $E=0$ put on the
whole of the system (\ref{rhodot2})-(\ref{C4}). In the same
viewpoint and by virtue of (\ref{Edot}) and (\ref{C3}), the time
derivative and the divergence of $E=0$ gives rise to two independent
constraints which are not preserved under evolution:
\begin{eqnarray}
    \label{C5}{\cal C}^5\equiv\textrm{curl}H-\frac{1}{2}\rho\s=0,\\
    \label{C3H}{\cal C}^{3,H}\equiv-\frac{1}{3}D\rho - [\s,H]=0.
\end{eqnarray}
It was not noted in \cite{Maartens2} that, although the \emph{sum}
equation ${\cal C}^3=0$ of $\textrm{div}\,E=0$ and ${\cal
C}^{3,H}=0$ is consistent under evolution, this is not necessarily
the case for both equations separately.  As $E=0$ has no influence
on ${\cal C}^1$ and ${\cal C}^2$, and as (\ref{C412}) is an
identity, the constraints (\ref{C1}), (\ref{C2}) and (\ref{C4})
\emph{do} evolve consistently as in the generic case.

The time derivative of (\ref{C5}), modulo (\ref{C1}) and (\ref{C5}),
yields a primary integrability condition ${\cal E}=0$ whose repeated
time differentiation leads to a non-terminating chain of new
constraints ${\cal C}_{H,1}$ after . The linearization of ${\cal E}$
around an FLRW background (\ref{FLRW}) reads $\frac{1}{6}\rho\t\s$,
such that linearized non-vacuum ID models with $E=0$ are either FLRW
or satisfy $\theta=0,\rho=2\Lambda$, as is seen from
(\ref{thetadot2}). Hence in the case $\Lambda=0$ \emph{no linearized
anti-Newtonian universes exist at all} \cite{Maartens2}. Also note
that, in the light of the question whether to incorporate a
cosmological constant or not, the same conclusion cannot be drawn
for general $\Lambda$).  However, for the purpose of proving
inconsistency in the full non-linear theory, the chain ${\cal
C}_{H,1}$ is even more unmanageable than its purely electric
analogue ${\cal C}_E$ (see Section 2.3).

Analogously, for the time derivative of ${\cal C}^{3,H}$ one finds
after a short calculation, on using (\ref{commTD}), (\ref{TSU}),
(\ref{sdot2}), (\ref{Hdot}) and (\ref{S2U}):
\begin{eqnarray}
    T({\cal C}^{3,H})=\left(\frac{1}{2}\s-\frac{5}{3}\t\right)\,{\cal C}^{3,H}+\frac{1}{3}{\cal J},\nonumber\\
    \label{J}{\cal J}=\rho D\theta +\left(\frac{1}{2}{\sigma}-\frac{1}{3}\theta\right) D\rho.
\end{eqnarray}
Henceforth,  ${\cal J}=0$. As $E\equiv 0$, and ${\cal C}^{3,H}=0$ is
the translation of $\textrm{div}\,E=0$ whereas ${\cal C}^5=0$ is
coming from $\dot{E}=0$, the same condition should be found on
calculating $\textrm{div}\,{\cal C}^5$ (see also (A13) in
\cite{Maartens1}). Indeed, in \cite{Maartens2} it was calculated
that \footnote{In \cite{Maartens2}, the $-\sigma{{\cal C}^{3}}$ term
was forgotten in the right hand side.}
\begin{eqnarray}\label{divC5}
    \textrm{div}\,{{\cal C}^{5}}&=&\frac{1}{2}\textrm{curl}\,{{\cal C}^{4}}
    - (\sigma+\frac{1}{3}\theta)\,{{\cal C}^{3}}-\frac{1}{2}{{\cal C}^{1}}-\frac{1}{3}{\cal J}.
\end{eqnarray}
The successive time derivatives of ${\cal J}^{(0)}:={\cal J}=0$ lead
to a second chain of constraints ${\cal C}_{H,2}=({\cal
J}^{(i)}=0)_{i=0}^\infty$, with ${{\cal J}^{(i)}}:=T^i({\cal J})$.
In \cite{Maartens2} ${{\cal J}^{(1)}}$ and ${{\cal J}^{(2)}}$ were
explicitly calculated (for $\Lambda=0$) and the general structure of
${{\cal J}^{(i)}}$ was given, but apparently the consequence of this
chain was overlooked. In Proposition 1 of Section 3, it will be shown
(for arbitrary $\Lambda$) that ${\cal C}_{H,2}$ implies $D\rho=0$.
But this gives rise to a \emph{third} chain ${\cal C}_{H,3}$ of
constraints, as follows. With $D\rho=0$, (\ref{C3H}) reads
$[\s,H]=0$. Taking the divergence of this and using (\ref{divSU}),
(\ref{C2}) and (\ref{C5}), one finds:
\begin{eqnarray}\label{K}
    {\cal K}\equiv\textrm{tr}(H^2)-\frac{1}{2}\rho\,\textrm{tr}(\s^2)=0,
\end{eqnarray}
and hence ${\cal C}_{H,3}=({\cal K}^{(i)}=0)_{i=0}^\infty$, with
${\cal K}^{(i)}:=T^i({\cal K})$\footnote{Notice that, without the
extra fact $D\rho=0$, the divergence of $\textrm{div}\,{\cal
C}^{3,H}$ already incorporates a Laplacian $D^2(\rho)$, which makes
the direct chain $T^i(\textrm{div}\,{\cal C}^{3,H})=0$ rather
unmanageable.}. In Theorem 1 of Section 3 it is finally shown that
within the class of ID's with $E=0$, ${\cal C}_{H,3}$ can only be
satisfied for $\s=H=0$, i.e., by the FLRW subclass.

\subsection{Comparison with purely electric ID}

Analogous to the purely magnetic case, (\ref{rhodot2})-(\ref{Edot})
with $H=0$ forms a `silent' system of evolution equations for the
remaining variables $\rho,\theta,\s,E$, but here (\ref{Edot})
is coupled to (\ref{rhodot2})-(\ref{sdot2}) via the $E$-term in
(\ref{sdot2}). Now $\dot{H}=0$ (together with $H=0$) and
$\textrm{div}\,H=0$ translate into
\begin{eqnarray}
    \label{C5E}{\cal C}^{5}&=&\textrm{curl}\,E=0,\\
    \label{C4E}{\cal C}^{4,E}&=&[\s,E]=0.
\end{eqnarray}
${\cal C}^1$ is independent of $H$, and ${\cal C}^2,{\cal C}^3$
contain $H$ only algebraically; hence the corresponding constraints
evolve consistently. In analogy with the $E=0$ case, the time
derivative of (\ref{C4E}), modulo the constraints
(\ref{C1})-(\ref{C3}) and (\ref{C5E}), gives rise to a new tensorial
condition ${\cal H}=0$ and an indefinite chain ${\cal
C}_E=(T^i({\cal H})=0)_{i=0}^{\infty}$ of constraints, identically
satisfied for Petrov type D and spatially homogeneous  models, but
not in general. However, in contrast to the magnetic case, the
linearizations of the $T^i({\cal H})$ around an FLRW background are
identically zero; hence Newtonian-like silent models are subject to
a \emph{linearization instability}, cf.\ Introduction.  The
difference is essentially due to the extra term
$\frac{1}{2}\rho\,\s$ in (\ref{C5}) within the magnetic case, not
present in (\ref{C5E}). The same absence makes that the divergence
of ${\cal C}^5$ doesn't lead to a new constraint in the electric
case:
\begin{eqnarray}
    \textrm{div}\,{{\cal C}^{5}}&=&\frac{1}{2}\textrm{curl}\,{{\cal C}^{3}}
     -(\sigma+\frac{1}{3}\theta)\,{{\cal C}^{4}}.
\end{eqnarray}
By the analogy of the remark made for the magnetic case, the
propagation of the constraint (\ref{C4E}) should not lead to a new
condition either. Indeed, this can easily be checked on using
(\ref{TSU}), (\ref{sdot2}), (\ref{Edot}) with $H=0$, and
(\ref{S2U}). Moreover, it was checked in \cite{Maartens4} that even
for general ID the condition $\textrm{div}\,H=0$, or equivalently
$[\s,E]=0$, is consistent under evolution, this making use of the
time derivative of (\ref{sdot2}) instead of using
(\ref{Edot})\footnote{Essentially, this comes down to showing
$[\s,T(\s)]=[\s,T^2(\s)]=0$. The first equation follows trivially
from (\ref{sdot2}) and $[\s,E]=0$; the second one then immediately
follows from time differentiation of the first and applying
(\ref{TSU}), but strangely enough the authors of \cite{Maartens4}
turned to a much longer but equivalent reasoning in an orthonormal
tetrad approach to show this.}. Hence there is no analogue here for
the chain ${\cal C}_{H,2}$ obtained in the magnetic case, and the
difference may also be traced back to the fact that, as in
electro-magnetism, $D\rho$ influences the divergence of the
electric, but not the magnetic field. Hence $\textrm{div}\,H=0$ does
not lead to direct information involving $D\rho$.

Finally notice that a counterpart for the third chain ${\cal
C}_{H,3}$ doesn't exist either: the divergence of (\ref{C4E}) leads
to an identity, since $\textrm{curl}\,E=0$ and since (\ref{C2}) now
reads $\textrm{curl}\s=0$!

Since $[\s,E]=0$, (\ref{rhodot2})-(\ref{Edot}) gives rise to a
autonomous dynamical system ${\cal S}$ in $\rho,\theta$, and two
Weyl and shear eigenvalues (or combinations thereof). Analyses for
$\Lambda=0$ in an orthonormal Weyl eigentetrad approach
\cite{vanElst} or coordinate approach \cite{Sopuerta} both led to
the same conclusion, which remains valid for general $\Lambda$: for
the spatially inhomogeneous Petrov type I subclass $C_{I,si}$ of the
purely electric silent models, the chain ${\cal C}_E$ eventually
translates into one or more indefinite chains of polynomial
constraints on ${\cal S}$. In \cite{Vandenbergh1}, four (compact)
relations on the variables ${\bf x}$ of ${\cal S}$ were found, by
which all these chains become identically satisfied. Hereby
$\Lambda$ seems to play the role of a bifurcation parameter for
${\cal S}$: for $\Lambda\leq 0$ and real ${\bf x}$, these relations
imply Petrov type 0 (FLRW), while for $\Lambda>0$ members of
$C_{I,si}$ emerge. Further analysis by computer (details of which will be given elsewhere) strongly indicates that these are the
\emph{only} members of $C_{I,si}$ for $\Lambda>0$, whereas there are
probably none for $\Lambda\leq 0$. This is a generalized statement
of the conjecture in \cite{vanElst,Sopuerta}. Still no definitive
proof has been found, mainly because of the massiveness of the
polynomials.

\section{Anti-Newtonian universes do not exist}

As outlined in Section 2.2, it is shown in this section that
non-vacuum irrotational dust models with $E=0$ imply $D\rho=0$
(Proposition 1) and hence are FLRW (Theorem 1), i.e., that
anti-Newtonian universes do not exist. Hereby, all statements and
expressions are implicitly assumed to be related to, resp.\ live on,
a small open subset $U$ of a spacetime. Since the whole of the
reasoning only involves a finite number of polynomial combinations
$F_i$ of tensorial quantities on $U$, one can always assume $U$
small enough such that for all couples $(F_i,F_j)$ either
$F_i(p)=F_j(p), \forall p \in U$ (denoted by $F_i=F_j$) or
$F_i(p)\neq F_j(p),\forall p \in U$ (denoted by $F_i\neq F_j$).

It should be remarked here that purely magnetic irrotational \emph{vacua}
($E=0,\omega=0,p=\rho=0$, with or without cosmological constant)
were shown to be inconsistent, first by Van den Bergh \cite{Vandenbergh2} in
an orthonormal tetrad approach, and later in a more transparent 1+3
covariant deduction (thereby generalizing the result from vacua to
spacetimes with vanishing Cotton tensor) by Ferrando and Saez
\cite{Ferrando}. \emph{Hence the vacuum case
$\rho=0$ may be excluded from the subsequent analysis}.\\

Some additional preparations should still be made. For the first step of the
proof of Proposition 1,  ${\cal
J}^{(i)}=T^i({\cal J})$ have to be calculated up to $i=6$. Denote $x_k$ for
$\textrm{tr}\,\s^k$ ($k\in\mathbb{N}$). By $(\ref{sdot2})$ with
$E=0$, $T(\s)$ is an element of the commutative (!) polynomial
algebra (over spacetime functions) generated by $\s$ itself, such
that $T(\s^k)=k\s^{k-1}T(\s)$. Taking the trace and substituting for
$T(\s)$ one finds
\begin{eqnarray}\label{Txk}
\label{trskdot}T(x_k)&=&k\left(-\frac{2}{3}\t\,x_k-x_{k+1}+\frac{1}{3}x_2 x_{k-1}\right).
\end{eqnarray}
Now the key remark is that \emph{all} of the $x_k$ for $k\geq 4$ are
polynomially dependent of $x_2$ and $x_3$ (according to the
formulation in \cite{Maartens2} w.r.t.\ these traces, this seems to
be exactly what was overlooked). Indeed, on using Newton's
identities up to power three, applied on the eigenvalues of $\s$,
the Cayley-Hamilton theorem for $\s$ reads
\begin{eqnarray}\label{CH}
    \s^{3}=\frac{1}{2}x_2\s+\frac{1}{3}x_3.
\end{eqnarray}
On multiplying (\ref{CH}) with $\s^k$ and taking the trace (or just
taking Newton's identities for powers higher than three) one gets
\begin{eqnarray}\label{xk}
    x_{k+3}=\frac{1}{2}x_2 x_{k+1}+\frac{1}{3}x_3 x_k,&\quad&k=1,2,\ldots.
\end{eqnarray}
Since $x_1=\textrm{tr}\,\s=0$ this reads $x_4=\frac{1}{2}{x_2}^2$ for $k=1$. Hence for $k=2,3$ (\ref{Txk}) becomes:
\begin{eqnarray}
\label{Tx2}T(x_2)&=&-\frac{4}{3}\t x_2-2x_{3}\\
\label{Tx3}T(x_3)&=&-2\t x_3-\frac{1}{2}x_{2}^2.
\end{eqnarray}
Hence (\ref{Tx2})-(\ref{Tx3}) together with
(\ref{rhodot2})-(\ref{sdot2}) forms an autonomous dynamical system
${\cal S}_H$ in the scalar invariants $\rho,\theta,x_2,x_3$.
Starting from (\ref{J}), it is easily seen that for the computation
of any ${\cal J}^{(i)}$, one only needs the basic operations
(\ref{TLeib}) and (\ref{commTD}), together with (\ref{sdot2}) and
the equations of ${\cal S}_H$ as substitution rules for
$T(\s),T(\rho),T(\theta),T(x_2)$ and $T(x_3)$. 
All ${\cal J}^{(i)}$ will be elements of the module generated by
$D\rho,\rho D\theta,\rho D x_2$ and $\rho D x_3$  over the
(commutative) polynomial ring
$R=\mathbb{Q}[\rho,\theta,x_2,x_3,\Lambda][s]$. In particular, one
finds for ${\cal J}^{(1)}$ and ${\cal J}^{(2)}$:
\begin{eqnarray*}
 \fl   {\cal J}^{(1)}=-\rho Dx_2-\left(\frac{3}{2}\s+\frac{5}{3}\t\right)\rho D\theta-\left(\frac{1}{3}(\rho+\Lambda)+\s^2+\frac{2}{3}\t\s-\frac{x_2}{2}-\frac{5}{9}\theta^2\right)D\rho,\\
 \fl  {\cal J}^{(2)}=2\rho D(x_3)+\left(\frac{5}{2}\s+\frac{13}{3}\theta\right)\rho D(x_2)+\left(\frac{7}{6}\rho-\frac{4}{3}\Lambda+4\s^2+\frac{19}{3}\t\s+2x_2+\frac{10}{3}\theta^2\right)\rho D(\theta)\\
\fl +\left((\frac{17}{12}\s+\frac{19}{18}\theta)\rho+(\frac{14}{9}\theta-\frac{1}{3}\s)\Lambda+3\s^3+4\theta\s^2-(\frac{x_2}{2}-\theta^2)\s-x_3-\frac{8}{3}\theta x_2-\frac{10}{9}\theta^3\right)D(\rho).
\end{eqnarray*}
Hence ${\cal J}={\cal J}^{(1)}={\cal J}^{(2)}=0$ can be solved for
$\rho D\theta,\rho D x_2$ and $\rho D x_3$, which expresses them as
elements of the module generated by $D\rho$ over $R$ \footnote{(\ref{Dx2h}) and (\ref{Dx3h}) were added for comparison with the
corresponding formulas in the case $\Lambda=0$ from
\cite{Maartens2}. Apparently there was a typo regarding the coefficient of
$x_2 \s D\rho$ in (\ref{Dx3h}).} 
\begin{eqnarray}
    \label{Dtheta}\rho D\theta&=&\left(\frac{1}{3}\theta-\frac{1}{2}{\sigma}\right) D\rho,\\
    \label{Dx2}\rho D x_2&=&-\left(\frac{1}{4}\s^2+\frac{1}{3}\theta \s-\frac{1}{2}x_2+\frac{1}{3}(\rho+\Lambda)\right)D\rho\\
    \label{Dx2h}&=&-\left(\frac{1}{4}\hat{\s^2}+\frac{1}{3}\theta \s-\frac{5}{12}x_2+\frac{1}{3}(\rho+\Lambda)\right)D\rho,\\
 \label{Dx3}\rho D x_3&=&\left(\Lambda(\frac{1}{4}\s+\frac{1}{6}\theta)-\frac{3}{16}\s^3-\frac{1}{8}\theta\s^2
+\frac{1}{8}x_2\s-\frac{1}{12}x_2\theta+\frac{1}{2}x_3\right)D\rho\\
\label{Dx3h}&=&\left(\Lambda(\frac{1}{4}\s+\frac{1}{6}\theta)-\frac{3}{16}\hat{\s^3}-\frac{1}{8}\theta\hat{\s^2}
+\frac{1}{8}x_2\s-\frac{1}{8}x_2\theta+\frac{7}{16}x_3\right)D\rho.
\end{eqnarray}

\vspace{.5cm}

The result of Proposition 1 is that anti-Newtonian universes
necessarily satisfy $D\rho=0$, and hence $[\s,H]=0$ from
(\ref{C3H}). As $\s$ and $H$ are both spatial and symmetric, this
implies that (a) $\s H$ is symmetric, such that one can write
(\ref{Hdot}) as
\begin{eqnarray}
    \label{Hdot2}T(H)=-\theta H+3\s H-\textrm{tr}(\s H),
\end{eqnarray}
and (b) that $\s$ and $H$ commute as (1,1)-tensors. Hence $T(\s)$
and $T(H)$ are elements of the commutative polynomial algebra (over
spacetime functions) generated by $\s$ and $H$, such that $T(\s^i
H^j)=\s^{i-1}H^{j-1}(i\,H\,T(\s)+j\,\s\, T(H))$ for arbitrary $i$
and $j$. Substituting for $T(\s),T(H)$ by
(\ref{sdot2}),(\ref{Hdot2}) and taking the trace yields:
\begin{eqnarray}\label{Txij}
    T(x_{i,j})&=&-(\frac{2i}{3}+j)\theta  x_{i,j}+(3j-i)x_{i+1,j}
          +\frac{i}{3}x_{2,0}x_{i-1,j}-jx_{1,1}x_{i,j-1},
\end{eqnarray}
wherein $x_{i,j}$ is written for $\textrm{tr}(\s^iH^j)$. On
multiplying (\ref{CH}) with $\s^{i}H^j$ and taking the trace these
are found to be related by
\begin{eqnarray}\label{CHij}
    x_{i+3,j}=\frac{1}{2}x_{2,0} x_{i+1,j}+\frac{1}{3}x_{3,0} x_{i,j},&\quad&i,j=0,1,2,\ldots.
\end{eqnarray}
For the first step of the proof of Theorem 1, 
${\cal K}^{(i)}=T^i(\cal K)$ needs to be computed up to $i=5$. Starting from ${\cal K}=
x_{0,2}-\frac{1}{2}\rho x_{2,0}$, inspection of (\ref{Txij}) for
$j=2$ learns that ${\cal K}^{(i)}, i=0\,..\,5$ will be of the form
$\frac{6!}{(6-i)!}x_{i,2}+$ other terms. Hence ${\cal K}^{(i)}=0$
determines $x_{i,2}$ as a polynomial in
$\rho,\theta,\Lambda,x_{j,2}(j<i)$ and some of the
$x_{k,0},x_{l,1}$. Substitution in a particular order of identities (\ref{CHij}) and obtained expressions  yields compact equations $u[i]=0,i=0\,..\,5$, where 
\begin{eqnarray}
  \label{u0}  u_0&=&x_{0,2}-\frac{1}{2}\rho x_{2,0}\\
    u_1&=&\frac{1}{6}x_{2,0}\rho\theta+6x_{1,2}+\rho x_{3,0}\\  u_2&=&-\frac{2}{3}x_{3,0}\rho\theta+\frac{1}{3}\rho {x_{2,0}}^2-\frac{1}{12}x_{2,0}\rho^2+\frac{1}{6}x_{2,0}\rho\Lambda+30x_{2,2}-12{x_{1,1}}^2\\
    u_3&=&\frac{1}{2}\rho^2x_{3,0}+\frac{1}{6}x_{2,0}\Lambda\rho\theta-2{x_{2,0}}^2\rho\theta+6x_{3,0}\rho x_{2,0}-108x_{1,1}x_{2,1}-x_{3,0}\rho\Lambda\\
    u_4&=&-18{x_{1,1}}^2x_{2,0}-{x_{2,0}}^3\rho+\frac{3}{4}{x_{2,0}}^2\rho^2-12\rho {x_{3,0}}^2-\frac{4}{3}x_{3,0}\Lambda\rho\theta+6x_{2,0}\rho\theta x_{3,0}\nonumber\\
&&+\frac{2}{9}\Lambda\rho\theta^2x_{2,0}-\frac{1}{12}x_{2,0}\rho^2\Lambda+\frac{1}{6}x_{2,0}\Lambda^2\rho-\frac{5}{3}{x_{2,0}}^2\rho\Lambda-216{x_{2,1}}^2\\    \label{u5} u_5&=&36{x_{1,1}}^2{x_{3,0}}-\frac{29}{18}{x_{2,0}}^2\Lambda\rho\theta-\frac{1}{3}{x_{2,0}}\Lambda\rho^2\theta-\frac{2}{3}{x_{2,0}}^3\rho\theta-36{x_{2,0}}{x_{1,1}}{x_{2,1}}\nonumber\\
&&-8{x_{3,0}}^2\rho\theta+6\rho{x_{2,0}}^2{x_{3,0}}-\frac{13}{2}{x_{2,0}}{x_{3,0}}\rho^2+\frac{5}{6}\rho^2\Lambda{x_{3,0}}-\frac{5}{3}{x_{3,0}}\Lambda^2\rho\nonumber\\
&&+\frac{10}{27}\Lambda{x_{2,0}}\theta^3\rho-\frac{20}{9}\Lambda\theta^2\rho{x_{3,0}}+15{x_{2,0}}{x_{3,0}}\rho\Lambda+\frac{5}{6}{x_{2,0}}\Lambda^2\rho\theta
\end{eqnarray}

{\bf Remark.} The calculation of ${\cal J}^{(i)},i=1\,..\,6$ and $u[i],i=\,0..\,5$ were preformed in the standard commutative environment of Maple (version 9.5), as allowed by the index-free notation and in the present situation. The used Maple code which reflects how the $u[i]$ were obtained, is explicitly given in Appendix A.\\  

The second step of the proofs of both Proposition 1 and Theorem 1
require two independent combinations of the non-zero shear
eigenvalues $\s_k$ and of the magnetic Weyl eigenvalues $H_k$ each
$(k=1,2,3)$. The following choice was made:
\begin{eqnarray}
    \label{chs}s_1:=-\frac{3}{2}\s_1, \quad t_1:=\frac{1}{2}(\s_2-\s_3),\\
    \label{chH}h_1:=-\frac{3}{2}H_1 \quad k_1:=\frac{1}{2}(H_2-H_3),
\end{eqnarray}
with inverse relations
\begin{eqnarray}
    \label{invchs}\s_1 = -\frac{2}{3}s_1,\quad \s_2 =\frac{1}{3}s_1+t_1,\quad \s_3 =\frac{1}{3}s_1-t_1,\\
    \label{invchH}H_1 = -\frac{2}{3}h_1,\quad H_2 =\frac{1}{3}h_1+k_1,\quad H_3 =\frac{1}{3}h_1-k_1.
\end{eqnarray}

\vspace{.2cm}

{\bf Lemma 1.} For a purely magnetic ID model (a) $s_1=0$ implies
FLRW and (b) $t_1=cs_1$, with $c$ a constant function, implies that
the shear tensor is degenerate.\\

{\bf Proof.} The shear tensor is a diagonalizable (1,1) tensor. Let
$P_k$ be the projector on any eigenvectorfield corresponding to the
eigenvalue $\s_k$. Then it is straightforward to deduce that
\begin{eqnarray}\label{Tlambda}
    T(\s_k)=\textrm{tr}(P_k\,T(\s))=-\frac{2}{3}\t\s_k-\sigma_k^2+\frac{1}{3}\textrm{tr}(\s^2).
\end{eqnarray}
where the second equation is valid for the specific situation (\ref{sdot2}). From this it follows that
\begin{eqnarray}
    \label{Ts1}T(s_1)&=&-\frac{2}{3}\t s_1+\frac{1}{3}s_1^2-t_1^2,\\
    \label{Tt1}T(t_1)&=&-\frac{2}{3}t_1(s_1+\t).
\end{eqnarray}
(a) If $s_1=0$ then also $t_1=0$ by (\ref{Ts1}), such that the ID
model is shear-free and hence FLRW. (b) The time derivative of
$t_1-cs_1=0$, substituting $t_1$ for $cs_1$, yields
$c(c-1)(c+1)s_1^2=0$. The case $s_1=0$ yields FLRW, whereas $c=0$ or
$c=\pm 1$ is exactly saying that the shear tensor is degenerate,
according to (\ref{invchs}).$\hfill\opensquare$\\

Finally, the following notations and properties concerning
multivariate polynomials are used. Consider the polynomial ring
$\mathbb{Q}[{\bf x}]$, where ${\bf x}$ stands for the $n$-tuple
$(x_1,\ldots,x_n)$. For two elements $p_1$ and $p_2$ of
$\mathbb{Q}[{\bf x}]$, $\textrm{gcd}(p_1,p_2)$ is written for one of
the two normalized forms of their greatest common divisor (i.e., for
which all coefficients are relatively prime integers); hereby
$\textrm{gcd}(p_1,p_2)\neq 1$, resp.\ $\textrm{gcd}(p_1,p_2)=1$,
indicates that $p_1$ and $p_2$ have, resp.\ do not have, a common
factor. Write $\textrm{Res}(p_1,p_2;x_i)$ for the resultant of $p_1$
and $p_2$ w.r.t.\ the variable $x_i$. It is well known that
$\textrm{Res}(p_1,p_2;x_i)$ is a polynomial over $\mathbb{Q}$ in the
variables $x_1,\ldots,x_{i-1},x_{i+1},\ldots,x_n$, which is an
element of the ideal of $Q[{\bf x}]$ generated by $p_1$ and $p_2$,
i.e., there exist polynomials $k_1,k_2\in Q[{\bf x}]$ for which
$\textrm{Res}(p_1,p_2;x_i)=k_1p_1+k_2p_2$. Hence, if ${\bf x}^0$ is
a root of $p_1$ and $p_2$ (i.e., a solution of $p_1({\bf
x})=p_2({\bf x})=0$) then
$(x_1^0,\ldots,x_{i-1}^0,x_{i+1}^0,\ldots,x_n^0)$ is a root of
$\textrm{Res}(p_1,p_2;x_i)$ (see e.g.\ \cite{CLO}). The following
properties are perhaps less known but sometimes help to avoid
cumbersome computations at the end of a proof by resultants:\\

{\bf Lemma 2.} (a) For $p_1,p_2\in \mathbb{Q}[{\bf x}]$ and any
$x_i$, $\textrm{Res}(p_1,p_2;x_i)$ is the zero polynomial if and
only if $\textrm{gcd}(p_1,p_2)\neq 1$.\\
 (b) When $n=2$, and if
$p_1(x_1,x_2)$ and $p_2(x_1,x_2)$ are homogeneous polynomials of
their arguments, then the system $\{p_1=0,p_2=0\}$
\begin{itemize}
    \item only has the trivial solution $(x_1,x_2)=(0,0)$ when $\textrm{gcd}(p_1,p_2)=1$;
    \item is equivalent with the equation $\gcd(p_1,p_2)=0$ when $\textrm{gcd}(p_1,p_2)\neq 1$.
\end{itemize}
{\bf Proof.} (a) see \cite{CLO}, p 158; (b) is an almost direct consequence of (a).$\hfill\opensquare$\\

With these preparations the main result may now be proved in two steps, as indicated.\\

{\bf Proposition 1}. An ID spacetime with $E=0$ satisfies $D\rho=0$.\\

{\bf Proof.} Calculating ${\cal J}^{(3)},\ldots,{\cal J}^{(6)}$,
substituting for $D\theta,Dx_2$ and $Dx_3$ by (\ref{Dtheta}),
(\ref{Dx2}) and (\ref{Dx3}) one gets four respective equations
$p_i\,D\rho=0$ with $p_i\in
R=\mathbb{Q}[\rho,\t,x_2,x_3,\Lambda][\s]$, $i=1\,..\,4$, i.e.,
$D\rho$ is in the kernel of the (1,1)-tensor $p_i$. In writing out
these equations in their components w.r.t.\ a shear eigenframe, one
gets twelve equations $p_{ij}D_j\rho=0$, $i=1\,..\,4,j=1\,..\,3$ (no
summation over $j$), where
$p_{ij}=p_{ij}(\rho,\theta,s_1,t_1,\Lambda)$ originates from $p_i$
by substituting the $j^{\textrm{th}}$ shear eigenvalue $\s_j$ for
$\s$, ${\s_1}^i+{\s_2}^i+{\s_3}^i$ for $x_i$ and finally the right
hand sides of (\ref{invchs}) for $\s_1,\s_2,\s_3$.

Suppose  $D\rho\neq0$. Then there exists a non-vanishing component
$D_j\rho$. Whitout loss of generality we may assume $D_1\rho\neq 0$.
Then $p_{i1}=0$ for $i=1\,..\,4$. One has e.g.:
\begin{eqnarray*}
\fl
p_{11}=-\frac{1}{3}\Lambda^2+(\frac{8}{9}{s_1}^2+2{t_1}^2+\frac{1}{6}\rho+\frac{4}{9}\theta
s_1-\frac{4}{9}\theta^2)\Lambda+(\frac{1}{3}{s_1}^2+\frac{7}{6}{t_1}^2)\rho\\ \hspace{-1.5cm}-4{s_1}^2{t_1}^2-4{s_1}\theta{t_1}^2-3{t_1}^4\\
\fl
p_{21}=((-\frac{23}{18}\theta+\frac{2}{9}{s_1})\rho+\frac{8}{3}{s_1}^3-20\theta{t_1}^2-\frac{68}{9}\theta{s_1}^2
-\frac{52}{9}\theta^2{s_1}+\frac{40}{9}\theta^3-4{t_1}^2{s_1})\Lambda\\
\hspace{-1.5cm}+(\frac{20}{9}\theta-\frac{4}{9}{s_1})\Lambda^2
+(-\frac{35}{9}\theta{s_1}^2+\frac{10}{9}{s_1}^3-\frac{245}{18}\theta{t_1}^2+\frac{26}{9}{t_1}^2{s_1})\rho\\
\hspace{-1.5cm}+16{t_1}^4{s_1}-\frac{16}{3}{s_1}^3{t_1}^2
+\frac{140}{3}\theta^2{s_1}{t_1}^2+\frac{124}{3}\theta{s_1}^2{t_1}^2+40\theta{t_1}^4.
\end{eqnarray*}
Now compute the resultants
$q_i(\theta,s_1,{t_1}^2,\Lambda):=\textrm{Res}(p_{11},p_{i1};\rho)$,
$i=2\,..\,4$ and next the resultants
$r_i(s_1,{t_1}^2,\Lambda):=\textrm{Res}(q_2,q_i;\theta),i=3,4$. One 
gets equations of the form
\begin{eqnarray}
r_3(s_1,{t_1}^2,\Lambda)&\equiv&\Lambda^2(3{t_1}^2-\Lambda)^4(2{s_1}^2+7{t_1}^2+\Lambda)^4r_3^*({s_1}^2,{t_1}^2,\Lambda)=0,\\   r_4(s_1,{t_1}^2,\Lambda)&\equiv&\Lambda^2(3{t_1}^2-\Lambda)^4(2{s_1}^2+7{t_1}^2+\Lambda)^4s_1r_4^*({s_1}^2,{t_1}^2,\Lambda)=0,
\end{eqnarray}
with $\textrm{gcd}(r_3^*,r_4^*)=1$.
If $s_1$ vanished then the spacetime would be FLRW according to
Lemma 1 (a), hence $D\rho=0$, which is a contradiction. Thus we
still have to investigate the following four cases:
\begin{enumerate}
   \item $\underline{\Lambda=0}$. One has $q_i(\theta,s_1,t_1,0)=t_1^2a_i(\theta,s_1,t_1)=0,i=2\,..\,4$,
   with $\textrm{gcd}(a_i,a_j)=1$ for $i\neq j$.
For $P_i({s_1},{t_1}^2)=\textrm{Res}(a_2,a_i;\theta),i=3,4$ one computes
that $\textrm{gcd}(P_3,P_4)={t_1}^2(7{t_1}^2+2{s_1}^2)^2(35{t_1}^4+42{t_1}^2{s_1}^2+16{s_1}^4)$.
If $t_1\neq 0$ then $s_1$ and $t_1$ must be solutions of $\textrm{gcd}(P_3,P_4)=0$ because of Lemma 2 (b);
if $t_1=0$ then $p_{11}(\rho,\theta,s_1,0,0)=\frac{1}{3}\rho {s_1}^2=0$. Thus both possibilities imply $\s=0$,
which leads to the contradiction $D\rho=0$, as above.
  \item $\underline{\Lambda=3{t_1}^2}$. One has
  $q_i(\theta,s_1,t_1,3{t_1}^2)=t_1^2(\theta+s_1)^2 b_i(\theta,s_1,t_1)=0,i=2\,..\,4$,
  with $\textrm{gcd}(b_i,b_j)=1$ for $i\neq j$. If $t_1=0$ then $\Lambda=0$, impossible since (i) has just been excluded.
  One has $p_{11}(\rho,-s_1,s_1,t_1,3{t_1}^2)=\frac{\rho}{3}\rho({s_1}^2+5{t_1}^2)=0$,
  such that $\theta+s_1=0$ again leads to $\s=0$ and contradiction.
  If $t_1\neq 0\neq \theta+s_1$ then $b_i(\theta,s_1,t_1)=0$, $i=2\,..\,4$.
  But for $Q_i(s_1,t_1)=\textrm{Res}(b_2,b_i;\theta),i=3,4$ one computes that $\textrm{gcd}(Q_3,Q_4)=1$.
  Hence $\s=0$ by Lemma 2 (b), contradiction.
  \item $\underline{\Lambda=-2{s_1}^2-7{t_1}^2}$. This case is exactly equivalent to the vanishing of the leading
coefficient of $p_{11}$, seen as a polynomial of $\rho$, i.e., one
has
$p_{11}(\rho,\theta,s_1,t_1,-2{s_1}^2-7{t_1}^2):=F(\theta,s_1,{t_1}^2)$,
the exact form of which is compact but not relevant. Set
$R_i(\rho,\theta,s_1,{t_1}^2)=p_{i1}(\rho,\theta,s_1,t_1,-2{s_1}^2-7{t_1}^2)$,
$i=2\,..\,4$, compute the resultants
$U_i(\theta,s_1,t_1)=\textrm{Res}(R_2,R_i;\rho)$ and then the
resultants $V_i(s_1,t_1)=\textrm{Res}(F,U_i;\theta)$, $i=3,4$. One
finds $\textrm{gcd}(V_3,V_4)=(7t_1^2+2s_1^2)^3(s_1^2+5t_1^2)^4$,
such that $\s=0$ via Lemma 2 (b), contradiction.
  \item $\underline{r_3^*=r_4^*=0}$. Then, according to Lemma 2 (a), ${s_1}^2$ and ${t_1}^2$ are
  necessarily solutions of the non-trivial homogeneous polynomial equation $\textrm{Res}(r_3^*,r_4^*;\Lambda)=0$.
Hence the ratio $(t_1/s_1)^2$ must be constant (the case $s_1=0$ is excluded), which can only be 1 or 0 by Lemma 1 (b).
Now one computes that
$\textrm{gcd}(r_3^*({s_1}^2,{s_1}^2,\Lambda),r_4^*({s_1}^2,{s_1}^2,\Lambda))=s_1^2(3{s_1}^2-\Lambda)$
and $\textrm{gcd}(r_3^*({s_1}^2,0,\Lambda),r_4^*({s_1}^2,0,\Lambda))=s_1^2\Lambda^3$,
which should be zero according to Lemma 2 (b); but this is impossible as (i) and (ii) have already been excluded.
\end{enumerate}
We conclude that the assumption $D\rho\neq 0$ was false and this finishes the proof.$\hfill\opensquare$\\

{\bf Theorem 1.} Any ID spacetime with $E=0$ is FLRW.\\

{\bf Proof.} From Proposition 1 and ${\cal C}^{4,H}=0$ one has
$[\s,H]=0$ and hence the chain ${\cal C}_{H,3}=({\cal K}^{(i)}\equiv
T^i({\cal K})=0)_{i=0}^\infty$, with ${\cal K}$ given by (\ref{K}).
From the above, the $u_i,i=0\,..\,5$ given by (\ref{u0})-(\ref{u5})
must vanish. As $x_{2,0}=0$ is equivalent with $\s=0$,
one may divide out factors which are powers of $x_{2,0}$ (or $\rho$)
in equations $F=0$. For any polynomial $p$ write $p^*$ for the
polynomial $p$ from which all such powers are taken out, and write
${\bf x}$ for the variable set
$(x_{2,2},x_{3,0},x_{2,1},x_{1,2},x_{2,0},x_{1,1},x_{0,2})$.
One subsequently computes $O_i({\bf
x},\rho,\theta)=\textrm{Res}^*(u_2,u_i;\Lambda)$, $P_i({\bf
x},\rho)=\textrm{Res}^*(u_1,O_i;\theta)$ and $Q_i({\bf
x})=\textrm{Res}^*(u_0,P_i;\rho)$ where $i$ runs from 3 to 5. Now in
the $Q_i$ perform the substitutions
$x_{i,j}={\s_1}^i{H_1}^j+{\s_2}^i{H_2}^j+{\s_3}^i{H_3}^j$ for the
index couples $(i,j)$ appearing in ${\bf x}$, followed by the
substitutions (\ref{invchs})-(\ref{invchH}). This yields polynomials
$R_i(s_1,t_1,h_1,k_1)$, $i=3\,..\,5$. The case $s_1=0$ leads to FLRW
by Lemma 1 (a), so we may assume $s_1\neq 0$ and scale $t_1$ with
$s_1$ and $h_1,k_1$ with $s_1^2$ to ease the symbolic computations.
This boils down to the substitution $s_1=1$ in the $R_i$. Now for
$Y_i({t_1}^2,{h_1}^2):=\textrm{Res}(R_1(1,t_1,h_1,k_1),R_i(1,t_1,h_1,k_1);k_1),i=2,3$
one computes that
$Z:=\textrm{gcd}(Y_1,Y_2)=(3{t_1}^6+75{t_1}^4-15{t_1}^2+1)(3{t_1}^2+1)^{29}{h_1}^{24}$.
The second factor is positive definite, and so is the first, as
$75{t_1}^4-15{t_1}^2+1=75(({t_1}^2-\frac{1}{10})^2+\frac{1}{3000})$.
This leaves two cases:
\begin{enumerate}
    \item $\underline{h_1=0}$. One has $R_i(1,t_1,0,k_1)={k_1}^4R_1^*({t_1}^2,{k_1}^2)$ a
    nd $R_2(1,t_1,0,k_1)={k_1}^6R_2^*({t_1}^2,{k_1}^2)$, where $\textrm{gcd}(R_1^*,R_2^*)=1$.
    Thus $\textrm{Res}(R_1^*,R_2^*;{k_1}^2)$ is a non-trivial polynomial in the scaled variable ${t_1}^2$ by Lemma 2 (a).
    Hence, if $k_1\neq 0$ then ${(t_1/s_1)}^2$ must be a constant, which can only be 1 or 0 by Lemma 1 (b).
But one has $R_1^*(1,1,{k_1}^2)\propto {k_1}^4(-5{k_1}^2+528)$, $R_1^*(1,1,{k_1}^2)\propto {k_1}^6(-31{k_1}^2+2544)$,
and $R_1^*(1,0,{k_1}^2)\propto {k_1}^4(-{k_1}^2+3)$, $R_2^*(1,0,{k_1}^2)\propto {k_1}^6({k_1}^2+3)$,
which leads to contradiction. If $k_1=0$ then $H=0$ and the spacetime is FLRW.
    \item $\underline{h_1\neq 0}$. Set $Z_i:=Y_i/Z,i=1,2$. Analogously as for $h_1=0$,
    $\textrm{Res}(Z_1,Z_2;h_1)$ is a non-trivial polynomial in the scaled variable ${t_1}^2$ by Lemma 2(a),
    and hence ${(t_1/s_1)}^2$ must be 1 or 0 by Lemma 1 (b). But one computes that
    $\textrm{gcd}(Z_1(1,{h_1}^2),Z_2(1,{h_1}^2))=\textrm{gcd}(Z_1(0,{h_1}^2),Z_2(0,{h_1}^2))=1$,
    i.e.\ $Z_1(1,{h_1}^2)$ and $Z_2(1,{h_1}^2)$, resp.\ $Z_1(0,{h_1}^2)$ and $Z_2(0,{h_1}^2)$,
    have no common solution, which leads to contradiction.
\end{enumerate}
All possible cases lead to FLRW or a contradiction, and this finishes the proof.$\hfill\opensquare$\\

\vspace{.2cm}

{\bf Remark.} The reason why the analysis was first performed in scalar invariants, and not
in the variables $s_1,t_1,h_1,k_1$ from the start, is twofold.
Firstly, in Proposition 1 the use of $D(x_2),D(x_3)$ instead of
$D(s_1),D(t_1)$ rules out the special case of degenerate shear (the
transformation between both sets has a Vandermonde-like
determinant). Secondly,  and w.r.t.\ the chain ${\cal C}_{H,3}$, 
the compactness and low degree of the
polynomials $u_i,i=0\,..\,5$ is striking: when expressed in $s_1,t_1,h_1,k_1$
the number of terms increases by a factor three to four. Although
eventually this is of no importance here, this may be kept in mind
for other consistency analyses, when one has to make a choice for
variables to be eliminated. 

\section{Conclusion and discussion}

It was proved in the full, nonlinear theory of general relativity that
purely magnetic irrotational dust (ID) models (`anti-Newtonian
universes') do not exist, irrespective as to whether a cosmological
constant $\Lambda$ is present or not. This was conjectured in
\cite{Maartens2} for $\Lambda=0$. The proof was mainly done in a 1+3
covariant approach, not making use of the Gauss equation for the
3–-Ricci curvature of 3–-surfaces orthogonal to $u$ (which yields
extra information within a tetrad formalism). The analysis for the
purely magnetic models was repeated for the purely electric ones,
and this summarized the main differences between both cases, as
partially pointed out in \cite{Maartens2}. The conjecture for
spatially inhomogeneous Petrov type I Newtonian-like ID models from
\cite{vanElst,Sopuerta} was properly restated in function of the
sign of $\Lambda$.

As a cosmological constant can be interpreted as a negative constant
pressure, ID spacetimes are related to the class of irrotational
models with an equation of state. In this respect, purely electric
or magnetic models for which the flowlines form a \emph{geodesic}
congruence are most closely related to the analysis here. This will
be dealt with in a forthcoming paper.

\ack LW is a research assistant supported by the Fund for Scientific
Research Flanders (F.W.O.), e-mail: lwyllema@cage.ugent.be. He wants
to thank N Van den Bergh for careful reading of the document and
suggestions, J Carminati for bringing the subject to the
attention, and K Vu and J Carminati for supplying useful computer algebra tools.

\begin{appendix}

\section{}

The following Maple code generates the expressions (\ref{u0})-(\ref{u5}): 

\begin{quotation}
    $x[4,0]:=1/2*x[2,0]^2$:\\
    for $i$ from 3 to 5 do\\
    $x[i,1]:=1/2*x[2,0]*x[i-2,1]+1/3*x[3,0]*x[i-3,1]$:\\
    $CHx2[i]:=x[i,2]=1/2*x[2,0]*x[i-2,2]+1/3*x[3,0]*x[i-3,2]$\\
    od:\\
    \\
    $K[0]:=x[0,2]-\frac{1}{2}*\rho*x[2,0]:u[0]:=K[0]$:\\
    verg:=${x[0,2]=\frac{1}{2}*\rho*x[2,0]}$:\\
    for $i$ to 5 do\\
 $K[i]:=$factor$(T(K[i-1]))$:\\
 $v[i]:=$factor(subs(verg,$K[i]$)):\\
 vgl[i]:=isolate$(v[i],x[i,2])$:\\
 if $i >= 3$ then $u[i]$:=factor(subs($CHx2[i]$,verg,$K[i]$)):\\
             else $u[i]:=v[i]$\\
 fi:\\
 verg:=verg union $\{$vgl[i]$\}$:\\
 od:
\end{quotation}
Herein, $T$ is a differential operator which acts on objects
$x_{i,j}$ according to (\ref{Txij}), which obeys
(\ref{rhodot2})-(\ref{sdot2}) with $E=0$ and
$\textrm{tr}\,\s^2=x_{2,0}$, and (\ref{Hdot2}) with $\textrm{tr}(\s
H)=x_{1,1}$. 

\end{appendix}

\end{document}